\documentstyle[preprint,aps,graphicx]{revtex}
\tightenlines
\begin{document}
\newcommand{\bref}[1]{eq.~(\ref{#1})}
\newcommand{\be}{\begin{equation}}
\newcommand{\en}{\end{equation}}
\newcommand{\bs}{$\backslash$}
\newcommand{\us}{$\_$}

\title{Roles of polarization, phase and amplitude in 
solid immersion lens systems}
\author{Lars Egil Helseth}
\address{University of Oslo, Department of Physics, N-0316 Oslo, Norway}

\maketitle
\begin{abstract}
By altering the polarization, phase and amplitude at the exit pupil, the 
intensity distribution near the focal plane of a Solid Immersion Lens(SIL) 
system can be changed. We have studied how the resolution and focal depth 
changes for a few particular cases. It was seen that by impinging radial 
polarization on a SIL system, we may obtain a rotational symmetric 
z-component of the focused wavefront with spot size similar to that predicted 
by scalar theory. We also observed that it was possible to manipulate 
the contributions from the homogeneous and inhomogeneous waves behind the SIL
by changing the amplitude and phase distribution at the aperture. In
this way it may be possible to improve both the resolution and focal depth
of the system. 
\end{abstract}

\narrowtext

\newpage

\section{Introduction}
The maximum resolution achievable with conventional optical techniques is
determined by the classical diffraction limit. The minimum optical spot 
diameter can be expressed as $\lambda/2NA$, where $\lambda$ is the wavelength 
in air, and $NA=nsin\alpha$ is the numerical aperture(n is the refractive 
index and $\alpha$ is the convergence semiangle). Fortunately, the diffraction 
limit can be circumvented by use of scanning near-field optical systems, where resolutions less than 50nm can be
achieved. Unfortunately, near-field techniques have been troubled by low
transmission efficiencies, and therefore poor signal to noise ratios. Although
recent research have improved the transmission efficiency considerably, it is
still only $1\%$ at 100nm spot size\cite{Minh}.  
Another way to increase the resolution is by application of a Solid Immersion
Lens(SIL)\cite{Ichimura,Milster}. This method is currently not capable of the same resolution as
the near-field techniques, but the light transmission efficiency is considerably 
better.  
The aim of the present paper is to describe how the polarization, phase and 
amplitude at the exit pupil influences the intensity distribution near the
focal plane of a SIL.  Although this is not a new
issue in optics(see ref.\cite{Ichimura,Milster,Stamnes} and references therein), 
we believe it is of interest to gain further understanding of SIL-systems as they 
may become an integral part of the future data storage systems.    
Since our focusing system has high NA and the focal point is placed many 
wavelengths away from the aperture, the diffracted field near the focal plane 
can be calculated using the Debye approximation\cite{Stamnes,Torok1}. 
Let us consider focusing through a dielectric interface(see fig. \ref{f1}). 
The electric field inside medium $i(=1,2)$ can be written 
as\cite{Stamnes,Torok1}:
\begin{equation}
\mbox{\boldmath $E$}_{i} =-\frac{ik_{i}}{2\pi}\int \int_{\Omega_{i}}
\mbox{\boldmath $T$} (\mbox{\boldmath $s$}_{i})
exp[ik_{i}(s_{ix}x+s_{iy}y+s_{iz}z)] ds_{ix}ds_{iy}
\label{a}
\end{equation}
Where $k_{i}=2\pi n_{i}/\lambda_{0}$ is the wavenumber, 
$\mbox{\boldmath$s$}_{i}=(s_{ix},s_{iy},s_{iz})$ is the unit vector along a 
typical ray, $\Omega_{i}$ is the solid angle formed by all the geometrical 
rays, $\mbox{\boldmath$T$}(\mbox{\boldmath $s$}_{i})$ is the vector pupil 
distribution, which accounts for the polarization, phase and amplitude 
distribution at the exit pupil. We can find the electric field near the focal
plane by matching the fields in the first and second medium at the
interface, $z=-d$. The resulting electric field in the second medium 
becomes\cite{Torok1}:
\begin{equation}
\mbox{\boldmath $E$}_{2} =C\int \int_{\Omega_{1}} \frac{\mbox{\boldmath $T$}
(s_{1x},s_{1y})}{s_{1z}}
exp[id(k_{2}s_{2z} -k_{1}s_{1z})]exp[ik_{2}s_{2z}z] exp[ik_{1}(s_{1x}x+s_{1y}y)] ds_{1x}ds_{1y}
\label{c}
\end{equation}
$C$ is a complex constant(which will be ignored in the rest of this paper). 
The unit wave-vector is defined in spherical coordinates:
\begin{equation}
\mbox{\boldmath $s$}_{i}=[sin(\theta_{i}) cos(\phi), sin(\theta_{i}) sin(\phi), cos(\theta_{i})]
\label{d}
\end{equation}
The position vector can be written as(see also fig. \ref{f1}):
\begin{equation}
\mbox{\boldmath $r$}_{c}=r_{c}[sin(\theta_{c}) cos(\phi_{c}), sin(\theta_{c}) sin(\phi_{c}),
cos(\theta_{c})]
\label{e}
\end{equation}
This gives the following diffraction integral:
\begin{equation}
\mbox{\boldmath $E$}_{2} =\int_{0}^{\alpha} \int_{0}^{2\pi} \mbox{\boldmath $T$}(\theta_{1}, \phi) 
exp[ik_{0}(r_{c} \kappa +\Psi)] sin(\theta_{1})d\theta_{1} d\phi
\label{f}
\end{equation}
\begin{equation}
\mbox{\boldmath $T$}(\theta_{1},\phi) = \mbox{\boldmath $P$}(\theta_{1},\phi) A(\theta_{1},\phi)
\label{g}
\end{equation}
where $\alpha$ is the convergence semiangle, $\mbox{\boldmath
$P$}(\theta_{1},\phi)$ is the polarization and $A(\theta_{1},\phi)$ represents  
the amplitude and phase distribution at the exit pupil. We will for the rest of
this paper assume that our optical system obeys the sine condition, 
$A(\theta_{1})\propto\sqrt{cos(\theta_{1})}$, see e.g. ref.\cite{Stamnes,Torok1}. 

\begin{equation} 
\kappa =n_{2}cos(\theta_{2}) cos(\theta_{c}) +n_{1}sin(\theta_{1})
sin(\theta_{c}) cos(\phi -\phi_{c})
\label{h}
\end{equation}
and 
\begin{equation}
\Psi =d[n_{2}cos(\theta_{2}) -n_{1}cos(\theta_{1})]
\label{i}
\end{equation}
$\Psi$ represents the aberration function introduced due to the 
mismatch in refractive index.
A detailed derivation of this integral was first presented
in reference\cite{Torok1}. We have presented it in a
slightly different form to enlight the further discussion in this
paper. We will consider two particular cases: 
1)Focusing of electromagnetic waves in a homogeneous media(air). In this case 
$n_{1}=n_{2}=1$. 2)Focusing of electromagnetic waves with a SIL, which means
that d=0(note that $n_{1}>n_{2}$). 
In both cases the aberration function $\Psi$ is identically zero\cite{Torok2}. 

\section{The influence of polarization}
The state of the polarization incident on the focusing system will influence
the resolution near the focal plane(see e.g. \cite{Stamnes,Sheppard}). 
To discuss this question quantitatively, we must find a general expression for the polarization vector.
We assume a incident polarization which may in general depend on 
the polar and azimuthal angle:
\begin{displaymath}
\mathbf{P_{0}} = 
\left[ \begin{array}{ccc} 
a(\theta_{1}, \phi) \\
b(\theta_{1}, \phi) \\
0                   \\ 
\end{array} \right]
\end{displaymath}
 
The polarization vector can be written as\cite{Torok1}:
\begin{equation}
\mathbf{P}(\theta_{1},\phi)=R^{-1}[L^{(2)}]^{-1}IL^{(1)}CRP_{0}
\end{equation}

\begin{displaymath}
\mathbf{R} = 
\left[ \begin{array}{ccc} 
cos(\phi) & sin(\phi) & 0 \\
-sin(\phi) & cos(\phi)& 0 \\
   0       & 0        & 1 \\
\end{array} \right]
\end{displaymath}

which describes the rotation of the co-ordinate system around the optical axis;

\begin{displaymath}
\mathbf{C} = 
\left[ \begin{array}{ccc} 
cos(\theta_{1}) & 0   &sin(\theta_{1})  \\
0               & 1        & 0 \\
-sin(\theta_{1}) &0    & cos(\theta_{1}) \\
\end{array} \right]
\end{displaymath}

which describes the change of polarization on propagation through the lens;

\begin{displaymath}
\mathbf{L^{(j)}} = 
\left[ \begin{array}{ccc} 
cos(\theta_{j}) & 0   &-sin(\theta_{j})  \\
0               & 1        & 0 \\
sin(\theta_{j}) &0    & cos(\theta_{j}) \\
\end{array} \right]
\end{displaymath}
 
which describes a rotation of the co-ordinate system into s and p-polarized
vectors; 

\begin{displaymath}
\mathbf{I} = 
\left[ \begin{array}{ccc} 
t_{p}        & 0               &0  \\
0            &t_{s}           & 0 \\
0            &0                &t_{p} \\
\end{array} \right]
\end{displaymath}
which represents the transmission(Fresnel coefficients) through the plane 
dielectric interface.

This results in:
\begin{displaymath}
\mathbf{P}(\theta_{1},\phi) = 
\left[ \begin{array}{ccc} 
a[t_{p}cos(\theta_{2})cos^{2}(\phi) +t_{s}sin^{2}(\phi)]
+b[t_{p}cos(\theta_{2})sin(\phi)cos(\phi)-t_{s}sin(\phi)cos(\phi)] \\
a[t_{p}cos(\theta_{2})cos(\phi)sin(\phi)-t_{s}sin(\phi)cos(\phi)] +
b[t_{p}cos(\theta_{2})sin^{2}(\phi)+t_{s}cos^{2}(\phi)] \\
- at_{p}sin(\theta_{2})cos(\phi)-bt_{p}sin(\theta_{2})sin(\phi)                \\ 
\end{array} \right]
\end{displaymath}    
In the special case of x-polarized incident light($a=1$ and b=$0$) this matrix 
reduces to\cite{Torok1};
\begin{displaymath}
\mathbf{P}(\theta_{1},\phi) = 
\left[ \begin{array}{ccc} 
\frac{1}{2}(t_{p}cos(\theta_{2}) +t_{s}) +\frac{1}{2}(t_{p}cos(\theta_{2})
-t_{s})cos(2\phi) \\
\frac{1}{2}(t_{p}cos(\theta_{2})-t_{s})sin(2\phi) \\
-t_{p}sin(\theta_{2})cos(\phi)\\ 
\end{array} \right]
\end{displaymath}
It is seen that different angles have different amplitudes, and that the
polarization is dependent on $\phi$. 
Another possibillity was explored by Quabis et. al. \cite{Quabis}, who found
that radial polarized light may increase the resolution. We extend their
analysis to a SIL system. We may write $a(\phi)=cos\phi$ and 
$b(\phi)=sin\phi$:

\begin{displaymath}
\mathbf{P}(\theta_{1},\phi) = 
\left[ \begin{array}{ccc} 
t_{p}cos(\theta_{2})cos(\phi)  \\
t_{p}cos(\theta_{2})sin(\phi) \\
-t_{p}sin(\theta_{2}) \\ 
\end{array} \right]
\end{displaymath}
When inserted in eq. (\ref{f}), this gives:
\begin{equation}
E_{2x}=iI_{1}^{rad}cos(\phi_{c})
\label{j}
\end{equation}
\begin{equation}
E_{2y}=iI_{1}^{rad}sin(\phi_{c}) 
\label{k}
\end{equation}
\begin{equation}
E_{2z}=-I_{0}^{rad} 
\label{l}
\end{equation}
\begin{equation}
I_{0}^{rad}=\int_{0}^{\alpha} B(\theta_{1})t_{p}sin(\theta_{2})
sin(\theta_{1})J_{0}(k_{1}r_{c}sin(\theta_{1})sin(\theta_{c}))exp(ik_{0}\Psi )exp(ik_{2}zcos(\theta_{2})) d\theta_{1}
\label{m}
\end{equation}
\begin{equation}
I_{1}^{rad}=\int_{0}^{\alpha} B(\theta_{1})t_{p}cos(\theta_{2})
sin(\theta_{1})
J_{1}(k_{1}r_{c}sin(\theta_{1})sin(\theta_{c}))exp(ik_{0}\Psi )exp(ik_{2}zcos(\theta_{2})) d\theta_{1}
\label{n}
\end{equation}
where $B(\theta_{1})=\sqrt{cos(\theta_{1})}$, and $J_{n}$ is the Bessel function
of the first kind, of order $n$.  
The z-component is completely independent of the azimuthal angle, and its
importance increases with increasing NA. 
Let us assume that we are able to find a medium which 
is only sensitive to the z-component of the polarization. That is, we may use 
the z-component, and disregard the x and y-components. Fig. \ref{f2} shows the energy density at the bottom
surface(the focal plane) of a SIL with $n_{1}=2$(and $n_{2}=1$), for both 
radial(solid line) and x-polarized incident light(the dashed line represents 
$\phi_{c}=0^{\circ}$, whereas the dotted line represents
$\phi_{c}=90^{\circ}$). In this example, we assume that
$\alpha=60^{\circ}$ and the wavelength $\lambda_{0}=635nm$. Fig. \ref{f2} 
indicates that it is possible to increase the resolution, and perhaps reproduce
the result predicted by scalar theory. 
However, it is worth noting that in practice one must combine the radial polarizer in
front of the focusing system with an annular aperture, both to increase
resolution and to avoid a singularity on the axis. This was discussed by Quabis
et. al.\cite{Quabis}. Therefore, if one could find a material which is sensitive
only to the z-component, a more detailed calculation using an annular aperture 
is required to determine the real resolution. 
  
\section{The influence of amplitude and phase}
The spatial amplitude and phase distribution(at the exit pupil) will also
influence the energy distribution near the focal plane(see e.g. 
\cite{Stamnes,Sheppard,Poon}). As with the incident polarization, there are countless ways to alter the phase
and transmittance. We will limit our discussion to the special class of amplitude/phase filters called
annular apertures(see ref.\cite{Ando,Yang} and references therein). Annular 
apertures that modifies the transmittance and/or phase distribution at the exit 
pupil may improve the resolution at the expense of higher sidelobes. In this 
section we will discuss the effect of such an aperture in front of a focusing 
system containing a SIL. The transmittance through an annular aperture can be 
expressed as:
\begin{displaymath}
A(\theta_{1}) =B(\theta_{1}) \left\{ \begin{array}{ll}
T_{1} & \textrm{if $0<\theta_{1}<\alpha_{1}$}\\
T_{2} & \textrm{if $\alpha_{1}<\theta_{1}<\alpha_{2}$}\\
\vdots & \\
T_{i} & \textrm{if $\alpha_{i-1}<\theta_{1}<\alpha_{i}$}\\ 
\vdots & \\
T_{N} & \textrm{if $\alpha_{N-1}<\theta_{1}<\alpha_{N}$}\\ 
\end{array} \right.
\end{displaymath}
where $T_{i}$ are complex constants(phase and amplitude) for the various zones in the aperture, and
$B(\theta_{1})=\sqrt{cos(\theta_{1})}$. If the incident light is polarized in the x-direction(a=1, b=0), the resulting 
field can be found by inserting the transmittance into eq. (\ref{f}):

\begin{equation}
E_{2x}=i[I_{0} +I_{2}cos(2\phi_{c})] 
\label{o}
\end{equation}
\begin{equation}
E_{2y}=iI_{2}sin(2\phi_{c}) 
\label{p}
\end{equation}
\begin{equation}
E_{2z}=2I_{1}cos(\phi_{c}) 
\label{q}
\end{equation} 
where 
\begin{equation}
I_{0}=\sum_{i=1}^{N} (T_{i}-T_{i+1}) \int_{0}^{\alpha_{i}}
A^{0x}(\theta_{1})J_{0}(k_{1}r_{c}sin(\theta_{1})sin(\theta_{c}))exp(ik_{0}\Psi )exp(ik_{2}zcos(\theta_{2}))d\theta_{1}
\label{r}
\end{equation}
\begin{equation}
I_{1}=\sum_{i=1}^{N} (T_{i}-T_{i+1}) \int_{0}^{\alpha_{i}} A^{1x}(\theta_{1})
J_{1}(k_{1}r_{c}sin(\theta_{1})sin(\theta_{c}))exp(ik_{0}\Psi ) exp(ik_{2}zcos(\theta_{2}))d\theta_{1}
\label{s}
\end{equation}
\begin{equation}
I_{2}= \sum_{i=1}^{N} (T_{i}-T_{i+1})\int_{0}^{\alpha_{i}} A^{2x}(\theta_{1})
J_{2}(k_{1}r_{c}sin(\theta_{1})sin(\theta_{c}))exp(ik_{0}\Psi )exp(ik_{2}zcos(\theta_{2}))d\theta_{1}
\label{t}
\end{equation}
\begin{equation}
A^{0x}=B(\theta_{1})(t_{s} +t_{p}cos(\theta_{2}))sin(\theta_{1})
\label{u}
\end{equation}
\begin{equation}
A^{1x}=B(\theta_{1})t_{p}sin(\theta_{2})sin(\theta_{1})
\label{v}
\end{equation}
\begin{equation}
A^{2x}=B(\theta_{1})(t_{s} -t_{p}cos(\theta_{2}))sin(\theta_{1})
\label{w}
\end{equation}
Where $T_{N+1}$ is defined to be zero, and we have used that 
$\int_{\alpha_{i}}^{\alpha_{i+1}}=\int_{0}^{\alpha_{i+1}}-\int_{0}^{\alpha_{i}}$
to derive these equations.
Similar expressions can be applied to the case of radial polarized light. 
From eqs. (\ref{r}), (\ref{s}) and (\ref{t}) we can see that the total field is 
a sum of electric fields from zones with increasing angular extent, and we
expect that the terms with the smallest $\alpha_{i}$ must be 
carefully balanced against each other in order to reduce the energy in the 
sidelobes. For high NA systems, the z-component becomes particularly important, and may
increase the sidelobe intensity substantially. Thus it is necessary to keep the
center peak ratio high in order to avoid a dominating z-component, which may
increase the the sidelobes as well as the spot size. 
In total one must not only balance the terms from the various zones, buth also
keep the sidelobes due to the z-(and y)component at an acceptable level. This is a 
difficult task, and is best treated by numerical analysis, e.g. by binary search 
methods\cite{Yang}. To keep the physics simple, we will limit ourselves to  
apertures with two and three zones. The profiles(the 
time-averaged energy density distributions) will be normalized, for simple
comparison between systems.

\subsection{Focusing in a homogenous media}
In order to compare with the situations occuring in a SIL-system, we first 
observe what happens when we place a three-zone aperture in front of a
system focusing in a homogeneous media. As pointed out by Ando\cite{Ando}, the 
center-peak intensity ratio can be maximized by using a phase aperture, and the 
sidelobes can be made small by maintaining the same phase for the light passing 
through the center and outer portion of the aperture.
On the basis of these results, let us assume that $T_{1}=1$, $T_{2}=-1$ and
$T_{3}=1$. Such an aperture can be produced by e.g. lithographical methods, and
was experimentally tested by Ando et. al.\cite{Ando1}. Let $\alpha_{1}=15^{\circ}$, $\alpha_{2}=30^{\circ}$, 
$\alpha_{3}=60^{\circ}$ and $\lambda_{0}=635nm$. 
Fig. \ref{f3} shows the focused beam profile for the three-zone phase aperture when 
$\phi_{c}=0^{\circ}$(solid line) and $\phi_{c}=90^{\circ}$(dash-dotted line).
For comparison we have also plotted the profile of the focused beam with no
aperture when $\phi_{c}=0^{\circ}$(dashed line) and
$\phi_{c}=90^{\circ}$(dotted line). 
The resolution due to the three-zone aperture is 
improved as compared to no aperture, but the peak sidelobe 
intensity is almost $20\%$. We could probably bring this number down by
increasing the number of zones in the annular aperture.
Fig. \ref{f4} shows the axial energy density distribution for the three-zone
annular aperture(solid line) and no annular aperture(dashed line). 
Note that the focal depth is larger with the three-zone aperture.  

\subsection{Focusing with a SIL}
The last ten years much research has been done on SIL systems\cite{Ichimura,Milster,Hasegawa,Guo,Shimura}. 
As mentioned in the introduction, a major goal has been
to improve the resolution and focal depth. To that end, an interesting 
possibility is the application of an annular aperture in 
front of a SIL. As have been pointed out by Milster $et$ $al$\cite{Milster}, 
plane waves incident on the bottom surface of the SIL will experience total 
reflection above the critical angle, $\theta_{c}=arcsin(n_{2}/n_{1})$. Thus we 
may divide the plane waves at the exit pupil into two parts; a homogeneous part 
where the angles are smaller than the critical angle at the interface, and an 
inhomogeneous part where the plane waves 
experience total reflection at the interface, corresponding to evanescent waves
below the bottom surface of the SIL. 
Let us first see what happens when we put a simple two-zone 
annular aperture with $T_{1}=0$ and $T_{2}=1$ in front of the SIL. 
The dotted lines in figs. \ref{f5} and \ref{f6} show the transverse 
profiles behind the bottom surface of the SIL($n_{1}=2.0$, $n_{2}=1.0$ and
$\lambda_{0}=635nm$) when 
$\phi_{c}=0^{\circ}$  and $\phi_{c}=90^{\circ}$, respectively. 
We assume that $\alpha_{1}=54^{\circ}$ and $\alpha_{2}=60^{\circ}$. 
For comparison, the profiles for a SIL with no
annular aperture are also shown(the dashed lines in figs. \ref{f5} and \ref{f6}). 
Note that as the center(dark) disk increases, the profile becomes more 
compressed, but the sidelobes increases. The increase in resolution
is more pronounced when $\phi_{c}=90^{\circ}$ than $\phi_{c}=0^{\circ}$, 
since the z-component of the electric field depends on
$cos\phi_{c}$(which  is zero when $\phi_{c}=90^{\circ}$, see eq. (\ref{q})). 
Fig. \ref{f7} shows the axial distribution behind the SIL with(dotted line) or
without(dashed line) a two-zone annular aperture. When the 
center disk increases, the focal depth decreases, contrary to what happens 
during focusing in a homogeneous media\cite{Poon}. This behaviour occur since the 
evanescent waves becomes more dominating when we block out the homogeneous
waves, and since the amplitude of evanescent waves are decreasing exponentially 
behind the SIL. To increase the light transmission efficiency, one may 
replace the two-zone amplitude aperture with an axicon, thus creating a bright 
ring.

Next we place a three-zone phase aperture(with the same properties as in the 
previous section) in front of the SIL. In
particular, we are interested in observing what happens when  
1)$\alpha_{1}<\alpha_{2}\leq\theta_{c}$ and 2)$\alpha_{2}>\alpha_{1}\geq\theta_{c}$.
The first case($\alpha_{1}<\alpha_{2}\leq\theta_{c}$) is in some ways similar to 
focusing in a homogeneous media. That is, we
expect the spot size to decrease, if the convergence semiangles have proper 
values. However, we also expect the focal depth to
decrease, since we balance terms with homogeneous waves against each other,
thus increasing the importance of the evanescent waves. As an example, let us
consider the same SIL-system as above, but now replace the two-zone aperture
with a three-zone annular aperture with  
$\alpha_{1}=20^{\circ}$, $\alpha_{2}=30^{\circ}$ and $\alpha_{3}=60^{\circ}$ .
The dash-dotted lines in
figs. \ref{f5} and \ref{f6} shows the resulting profiles when  
$\phi_{c}=0^{\circ}$ and $\phi_{c}=90^{\circ}$, respectively.
Note that the profile is slightly smaller than the dashed line(no annular 
aperture) when $\phi_{c}=90^{\circ}$, whereas for $\phi_{c}=0^{\circ}$ it
is slightly larger.
The dash-dotted line in fig. \ref{f7} shows the axial distribution. As expected,
the focal depth is smaller with the annular aperture. 

In the second case($\alpha_{2}>\alpha_{1}\geq \theta_{c}$) the opposite
behaviour may take place. That is, we are able to balance terms with 
inhomogeneous waves against each other, thus increasing 
the importance of the homogeneous waves. In this way we can produce some kind 
of enhancement of the focal depth, and if the the angles $\alpha_{1}$ and $\alpha_{2}$ are properly
chosen, the beam profile may become narrower as well. As an example we consider an
annular aperture with $\alpha_{1}=31^{\circ}$ and $\alpha_{2}=43^{\circ}$. 
The solid lines in 
figs. \ref{f5} and \ref{f6} are representing the transverse profiles
when $\phi_{c}=0^{\circ}$ and $\phi_{c}=90^{\circ}$, respectively. Now the 
situation is almost opposite compared with the previous annular aperture. That 
is, the profile is smaller than the dashed line(no aperture) when $\phi_{c}=0^{\circ}$, 
but larger when $\phi_{c}=90^{\circ}$.
The axial distribution(solid line in fig. \ref{f7}) is decaying more slowly than 
with no annular aperture, which confirms the discussion above.

\section{conclusion}
The choice of incident  
polarization, phase and amplitude may change the intensity 
distribution near the focal plane of a focusing system. 
It was seen that by impinging radial polarization on a SIL system, we 
may obtain a rotational symmetric z-component of the focused wavefront with spot
size similar to that predicted by scalar theory.  
We also discussed how annular apertures may change the resolution and focal depth. 
It was shown that when we place a three-zone phase aperture in front of a 
focusing system, both the resolution and focal depth may increase. 
When applied to SIL-systems, we observed that it was possible to manipulate 
the contributions from the homogeneous and inhomogeneous waves behind the SIL
by changing the phase and transmittance distribution at the aperture. In
this way it may be possible to improve both the resolution and focal depth. Future investigations may be directed towards the use of more 
than three zones to improve the resolution and focal depth.
We would like to point out that only a few special cases were discussed in this
paper, and that our examples were not optimized for any particular
applications.

\acknowledgements
I would like to thank all the people working in the
Superconductivity and Magnetooptics group at the University of Oslo for creating a 
very active and inspiring enviroment, which resulted in the ideas presented in 
this work. The research was financially supported by The Norwegian Research 
Council.

\newpage

\begin{figure}
\caption{The general geometry for focusing through a planar dielectric 
interface, located at z=-d. The focal plane is located at z=0.  
\label{f1}}
\end{figure}

\begin{figure}
\caption{The transversal time-averaged electric energy density when 
$n_{1}=2$, $n_{2}=1$ and $\alpha=60^{\circ}$. 
1)z-component of the radial polarized focused beam.
2)Total energy density for incident x-polarized light(a=1 and b=0) when 
$\phi_{c}=0^{\circ}$. 
3)Total energy density for incident x-polarized light(a=1 and b=0) when 
$\phi_{c}=90^{\circ}$.
\label{f2}}
\end{figure}

\begin{figure}
\caption{The transversal time-averaged electric energy density with and without
a three-zone annular aperture($\alpha_{1}=15^{\circ}$ and
$\alpha_{2}=30^{\circ}$) when $n_{1}=n_{2}=1$ and 
$\alpha=60^{\circ}$. 
1)Three-zone annular aperture when $\phi_{c}=0^{\circ}$
2)No annular aperture when $\phi_{c}=0^{\circ}$.
3)No annular aperture when $\phi_{c}=90^{\circ}$.
4)Three-zone annular aperture when $\phi_{c}=90^{\circ}$    
\label{f3}}
\end{figure}

\begin{figure}
\caption{The axial time-averaged electric energy density when  
$n_{1}=n_{2}=1$ and $\alpha=\alpha_{3}=60^{\circ}$. 
1)Three-zone annular aperture with $\alpha_{1}=15^{\circ}$ and
$\alpha_{2}=30^{\circ}$.
2)No annular aperture. 
\label{f4}}
\end{figure}

\begin{figure}
\caption{The transversal time-averaged electric energy density when
$\phi_{c}=0^{\circ}$, $n_{1}=2$, $n_{2}=1$ and $\alpha=\alpha_{3}=60^{\circ}$. 
1)Three-zone phase aperture with $\alpha_{1}=31^{\circ}$ and 
$\alpha_{2}=43^{\circ}$.
2)No annular aperture.
3)Two-zone annular aperture with $\alpha_{1}=54^{\circ}$.
4)Three-zone aperture with $\alpha_{1}=20^{\circ}$ and 
$\alpha_{2}=30^{\circ}$.  
\label{f5}}
\end{figure}

\begin{figure}
\caption{The transversal time-averaged electric energy density when
$\phi_{c}=90^{\circ}$, $n_{1}=2$, $n_{2}=1$ and $\alpha=\alpha_{3}=60^{\circ}$. 
1)Three-zone phase aperture with $\alpha_{1}=31^{\circ}$ and 
$\alpha_{2}=43^{\circ}$.
2)No annular aperture.
3)Two-zone annular aperture with $\alpha_{1}=54^{\circ}$.
4)Three-zone aperture with $\alpha_{1}=20^{\circ}$ and 
$\alpha_{2}=30^{\circ}$.   
\label{f6}}
\end{figure}

\begin{figure}
\caption{The axial time-averaged electric energy density when
$n_{1}=2$, $n_{2}=1$ and $\alpha=\alpha_{3}=60^{\circ}$. 
1)Three-zone phase aperture with $\alpha_{1}=31^{\circ}$ and 
$\alpha_{2}=43^{\circ}$.
2)No annular aperture.
3)Two-zone annular aperture with $\alpha_{1}=54^{\circ}$.
4)Three-zone aperture with $\alpha_{1}=20^{\circ}$ and 
$\alpha_{2}=30^{\circ}$.
\label{f7}}
\end{figure}

\newpage
\centerline{\includegraphics[width=14cm]{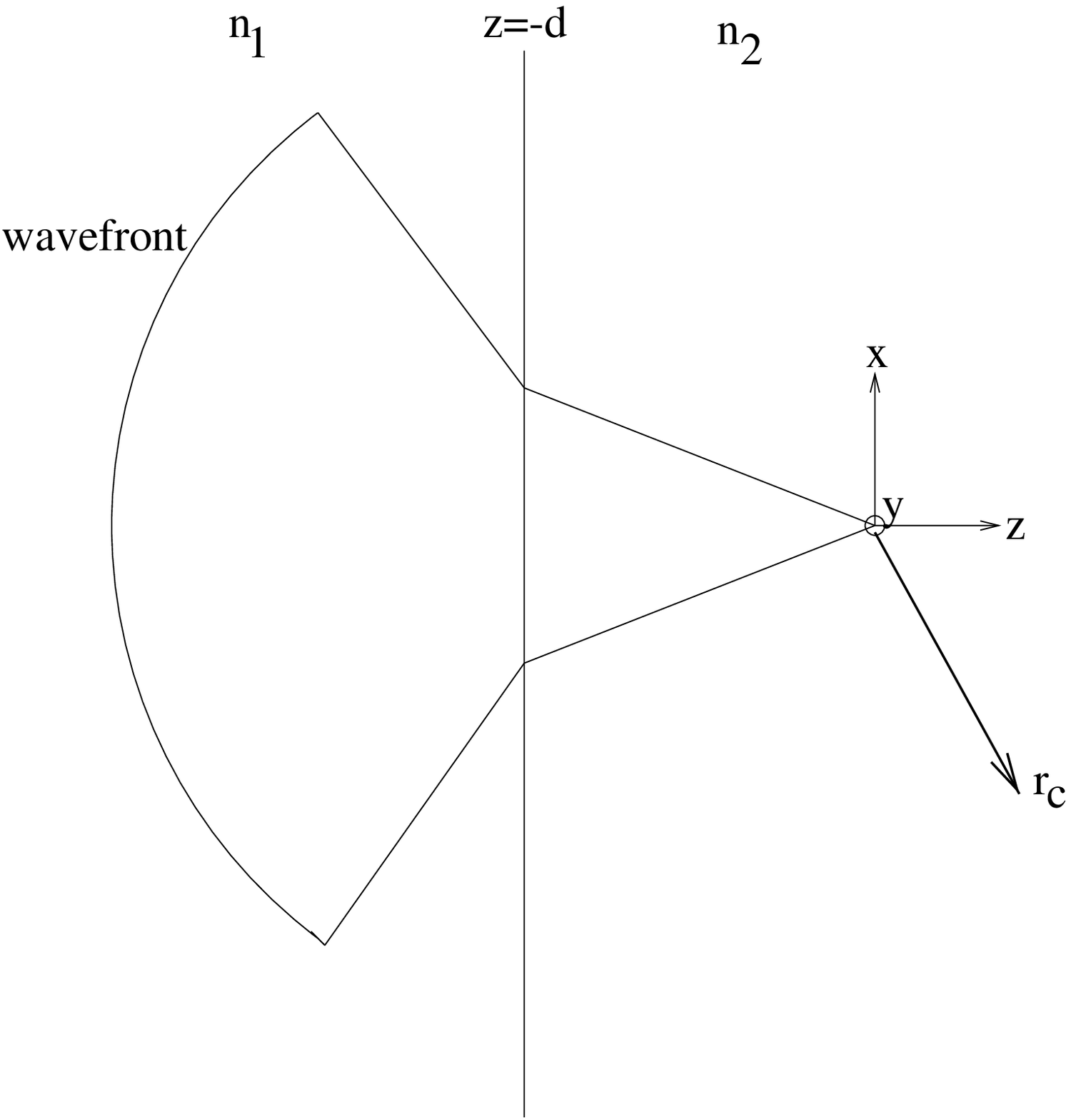}}
\vspace{2cm}
\centerline{Figure~\ref{f1}}

\newpage
\centerline{\includegraphics[width=14cm]{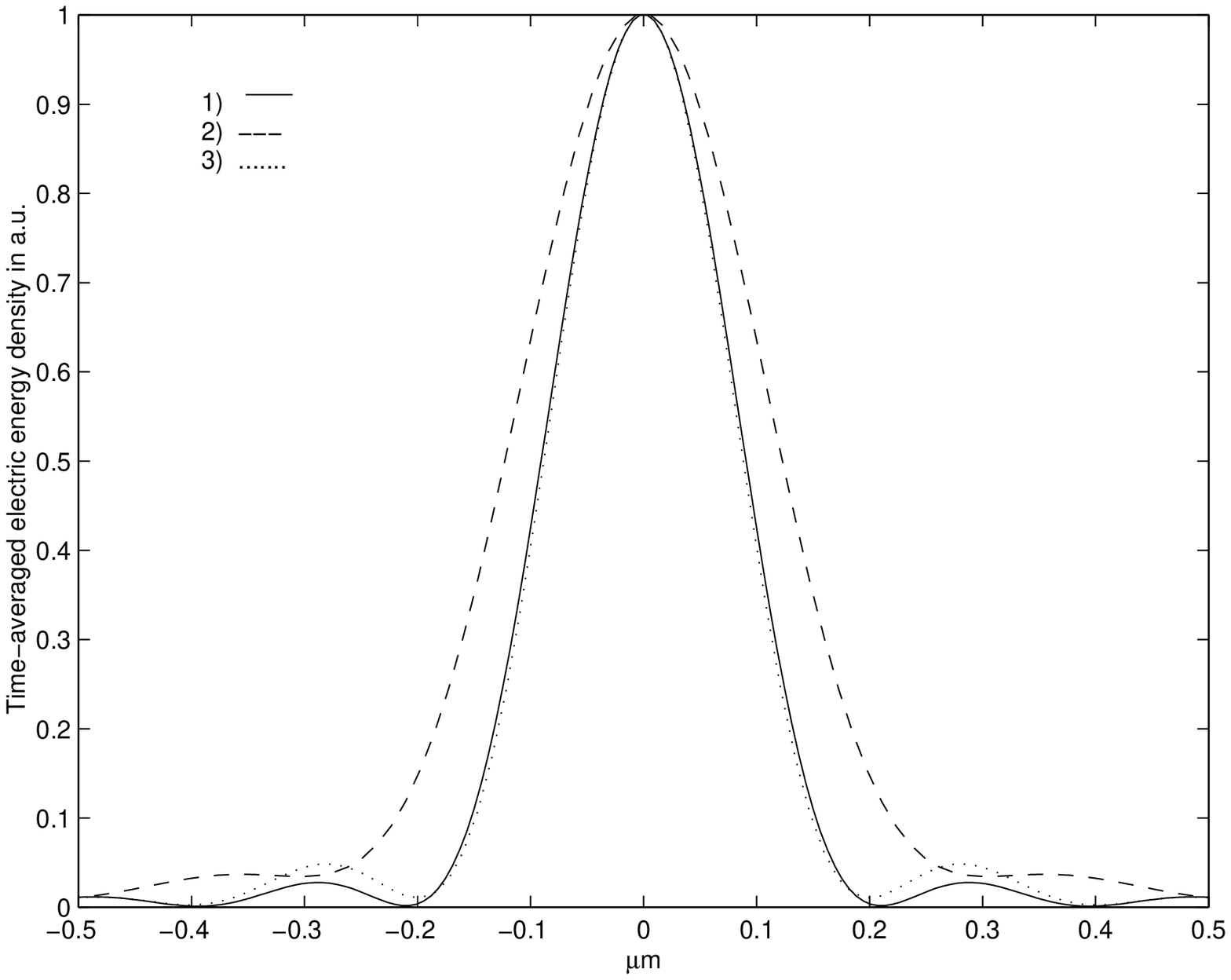}}
\vspace{2cm}
\centerline{Figure~\ref{f2}}

\newpage
\centerline{\includegraphics[width=14cm]{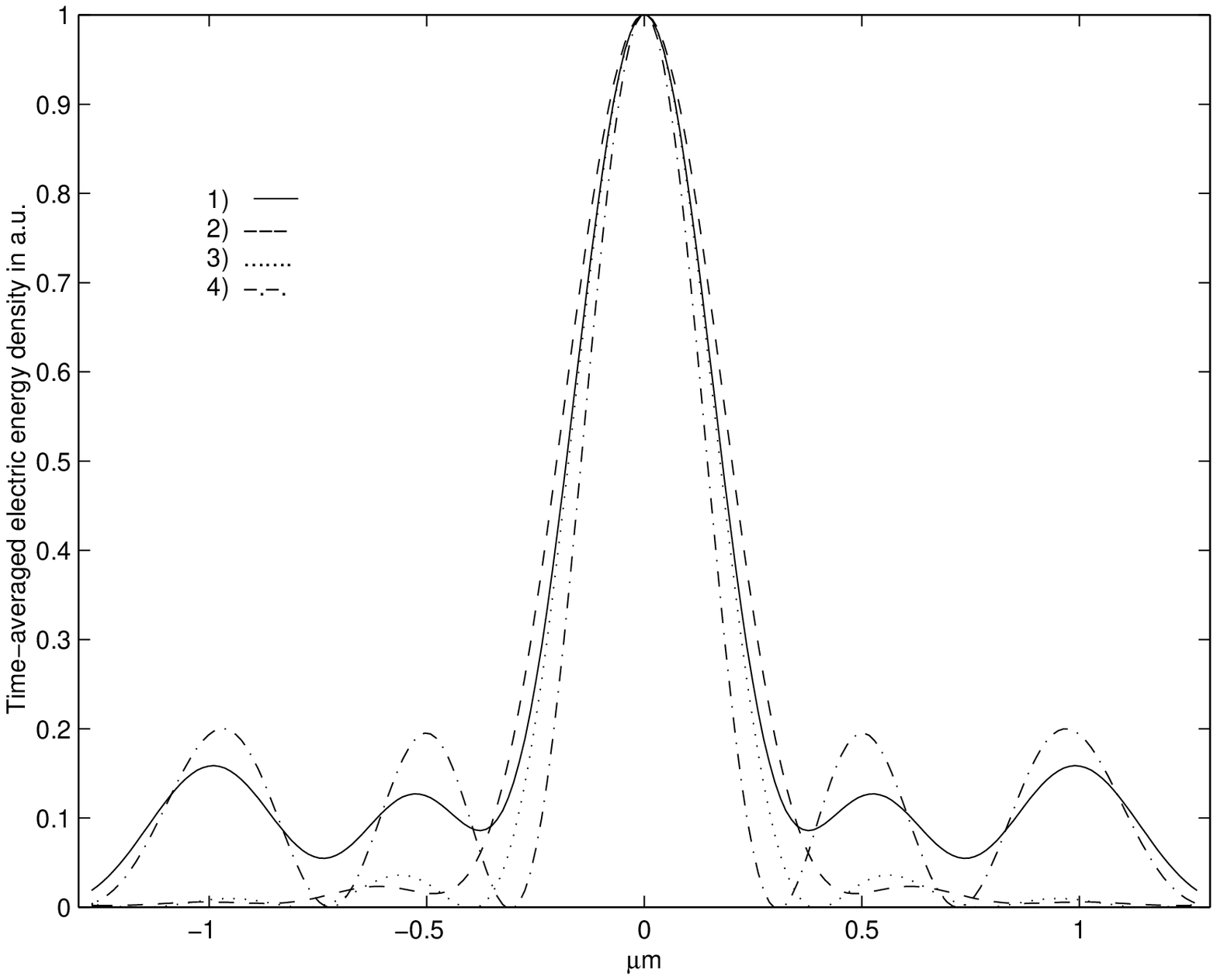}}
\vspace{2cm}
\centerline{Figure~\ref{f3}}

\newpage
\centerline{\includegraphics[width=14cm]{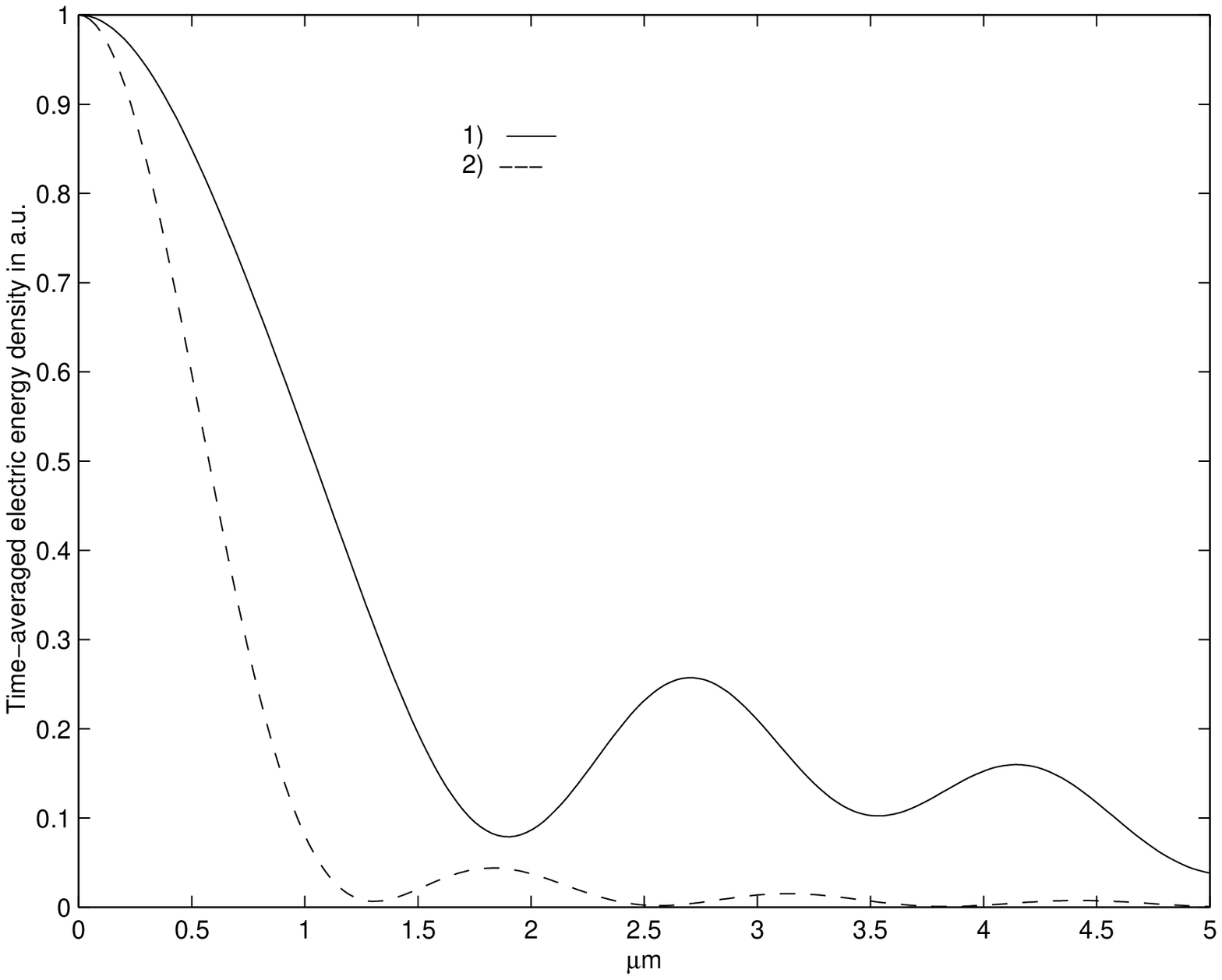}}
\vspace{2cm}
\centerline{Figure~\ref{f4}}

\newpage
\centerline{\includegraphics[width=14cm]{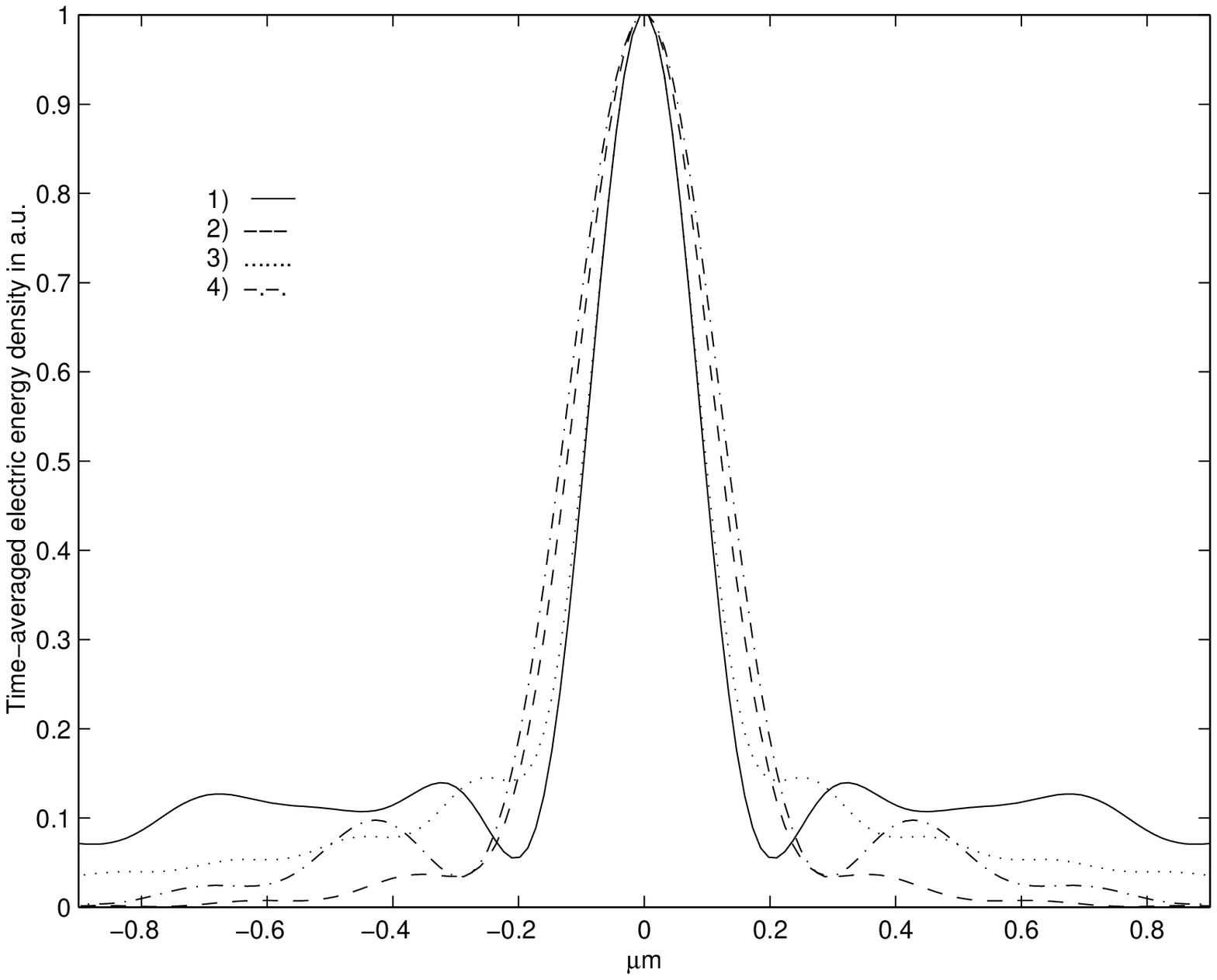}}
\vspace{2cm}
\centerline{Figure~\ref{f5}}

\newpage
\centerline{\includegraphics[width=14cm]{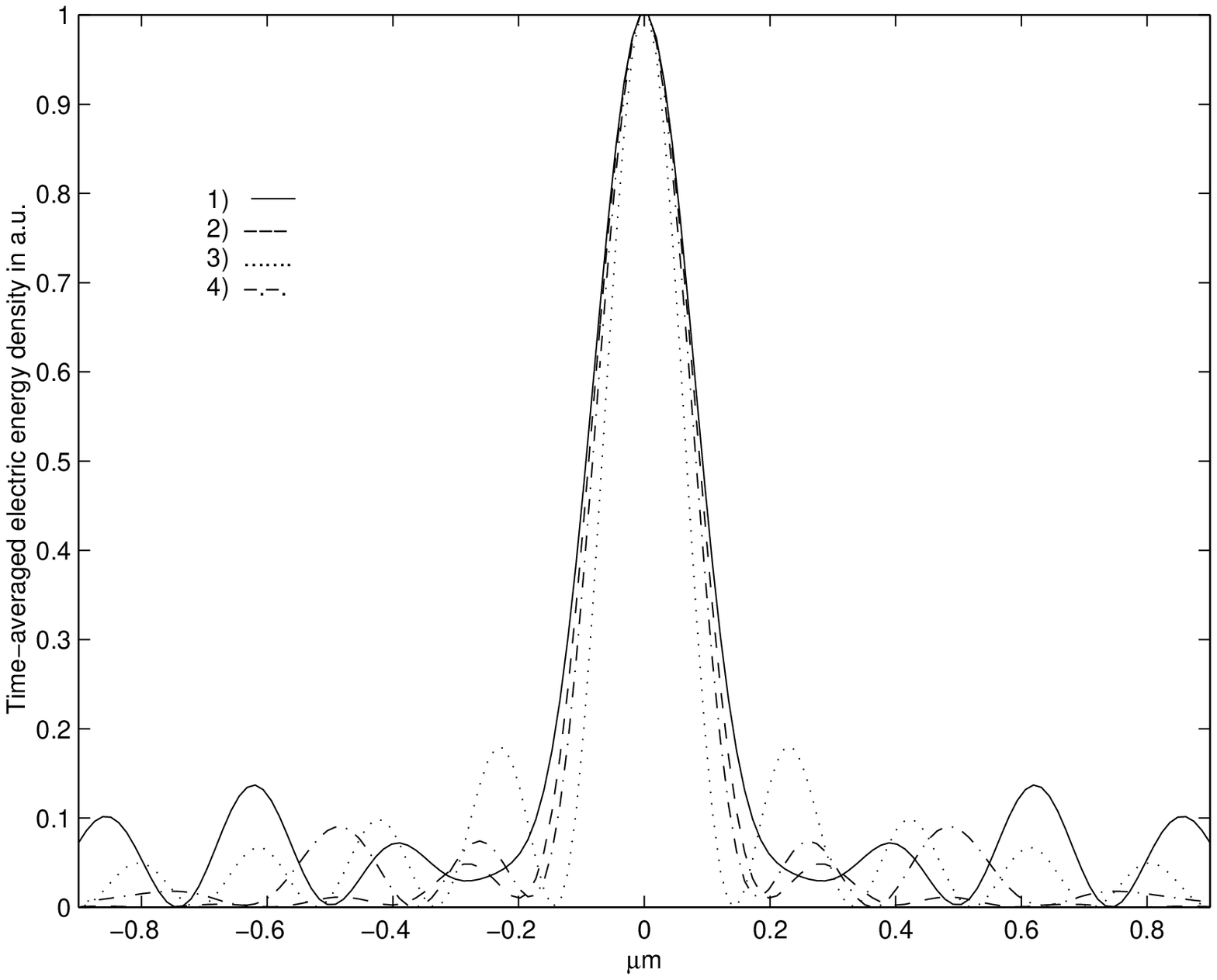}}
\vspace{2cm}
\centerline{Figure~\ref{f6}}

\newpage
\centerline{\includegraphics[width=14cm]{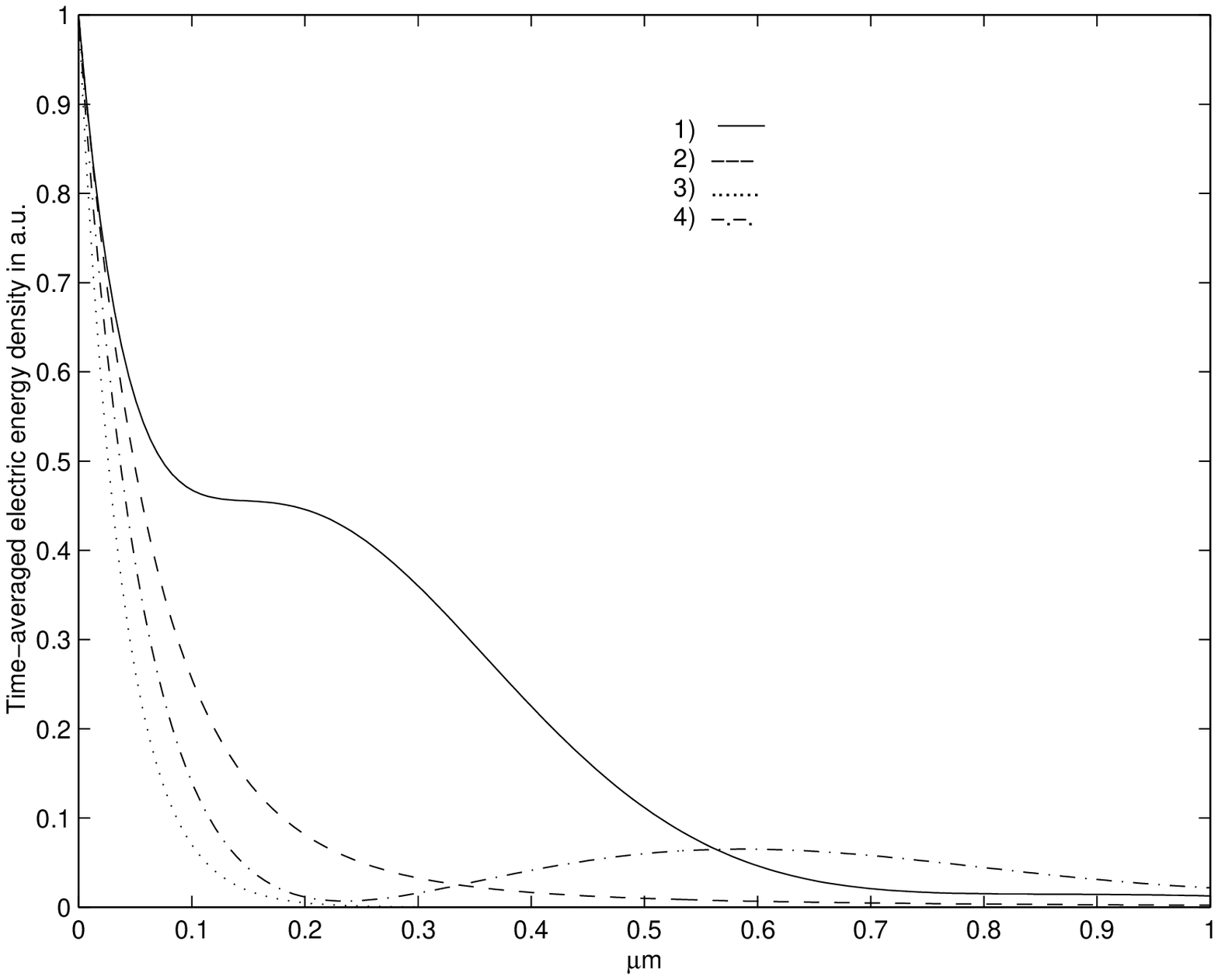}}
\vspace{2cm}
\centerline{Figure~\ref{f7}}

\end{document}